\documentclass[sigplan,screen]{acmart}
\usepackage[utf8x]{inputenc}

\pdfoutput=1

\usepackage{color}
\definecolor{keywordcolor}{rgb}{0.7, 0.1, 0.1}   % red
\definecolor{commentcolor}{rgb}{0.4, 0.4, 0.4}   % grey
\definecolor{symbolcolor}{rgb}{0.0, 0.1, 0.6}    % blue
\definecolor{sortcolor}{rgb}{0.1, 0.5, 0.1}      % green

\usepackage{listings}

\usepackage[shortlabels]{enumitem}
\usepackage{tikz}

\input{quiver.tex}

\newtheorem*{theorem*}{Theorem}

\author{Kexing Ying}
\affiliation{%
   \institution{University of Cambridge}
   \city{Cambridge}
   \country{UK}}
\email{ky316@cam.ac.uk}

\author{Rémy Degenne}
\affiliation{%
   \institution{Univ. Lille, Inria, CNRS, Centrale Lille, UMR 9198-CRIStAL}
   \city{Lille}
   \country{France}}
\email{remy.degenne@inria.fr}

\title{A Formalization of Doob's Martingale Convergence Theorems in mathlib}



\setcopyright{licensedothergov}
\acmPrice{15.00}
\acmDOI{10.1145/3573105.3575675}
\acmYear{2023}
\copyrightyear{2023}
\acmSubmissionID{poplws23cppmain-p22-p}
\acmISBN{979-8-4007-0026-2/23/01}
\acmConference[CPP '23]{Proceedings of the 12th ACM SIGPLAN International Conference on Certified Programs and Proofs}{January 16--17, 2023}{Boston, MA, USA}
\acmBooktitle{Proceedings of the 12th ACM SIGPLAN International Conference on Certified Programs and Proofs (CPP '23), January 16--17, 2023, Boston, MA, USA}
\received{2022-09-21}
\received[accepted]{2022-11-21}

\begin{document}

\begin{abstract}
We present the formalization of Doob's martingale convergence theorems in the mathlib library for the Lean theorem prover.
These theorems give conditions under which (sub)martingales converge, almost everywhere or in $L^1$.
In order to formalize those results, we build a definition of the conditional expectation in Banach spaces and develop the theory of stochastic processes, stopping times and martingales.
As an application of the convergence theorems, we also present the formalization of Lévy's generalized Borel-Cantelli lemma.
This work on martingale theory is one of the first developments of probability theory in mathlib, and it builds upon diverse parts of that library such as topology, analysis and most importantly measure theory.
\end{abstract}

\maketitle{}

\section{Introduction}
\label{sec:introduction}
% !TeX root = ../martingales_in_lean.tex

Martingales are a fundamental object of study in modern probability theory. First introduced by 
Lévy in 1934, martingale theory in the mathematical sense was developed by Doob in the 1940's.
Today, martingales have formed the foundations of stochastic calculus and have 
found applications in economics, finance and statistics among many other fields. 

Informally, a martingale models a series of fair games, while the related sub/super-martingales model the outcomes of a series of games biased towards/against the player.
The martingale convergence theorems are powerful result in martingale theory. Developed by Doob, they show that a sub/super-martingale converges 
provided some boundedness condition. It has a wide range of consequences, most notably 
the strong law of large numbers, the Radon-Nikodym theorem and Lévy's generalized Borel-Cantelli 
lemma.

We describe our formalization of martingale theory in the Lean theorem prover and provide 
the formalization of the almost everywhere martingale convergence 
theorem and the \(L^1\) martingale convergence theorem. To achieve this, we use and develop Lean's mathematical 
library \lstinline{mathlib} \cite{mathlib}. This entire project consists of approximately 9200 lines of code, all of which 
has been merged into \lstinline{mathlib}. The source code for this project 
can be found in the \lstinline{mathlib} GitHub repository\footnote{\href{https://github.com/leanprover-community/mathlib}{https://github.com/leanprover-community/mathlib}}.
\lstinline{mathlib} is a constantly evolving library: this paper discusses the code as it was at the time of the conference.

As an overview, this project involved the development and formalization of 
\begin{itemize}
  \item the definitions of conditional expectation, stopping time, hitting time and martingale;
  \item theory of uniform integrability and the Vitali convergence theorem;
  \item the optional stopping theorem/fair game theorem;
  \item Doob's maximal inequality;
  \item the a.e. and \(L^1\) martingale convergence theorems;
  \item Lévy's upward theorem;
  \item Doob's decomposition theorem;
  \item the one-sided martingale bound and
  \item Lévy's generalized Borel-Cantelli lemma.
\end{itemize}

A proof of the almost everywhere martingale convergence theorem has previously been formalized 
by \cite{vajjha2022formalization} earlier this year for the proof of the Dvoretzky’s stochastic approximation theorem in Coq.  
However, in contrast to their formalization, this project provides a more general framework to work 
with probability theory overall. In particular, our project contains a definition of 
the conditional expectation which applies to general Banach space-valued random variables, and thus, allowing
us to define Banach space-valued martingales instead of only real-valued ones. This generality is useful and is 
required for the theory of UMD-spaces (c.f. \cite{hytonen2016analysis}). 
Furthermore, this project provides the first formalization of the \(L^1\) martingale convergence theorem. 

Other than the aforementioned project, \cite{isabelle} defines real-valued martingales in Isabelle/HOL for the purpose of 
formalizing the Cox-Rubinstein model in financial mathematics. Similarly, the HOL4 library \cite{hol4} also formalizes 
the definition of the martingale property however, does so without showing the existence of conditional expectations. 
These two libraries do not contain any significant results on martingales beyond their definition. 

More broadly, parts of probability theory have been formalized in many systems. Most notably, the Isabelle/HOL library 
contains a general formulation of measure-theoretic probability theory including a proof of the central limit 
theorem \cite{avigad2017formally}, the laws of large numbers \cite{lln2021isabelle} and several definitions central 
to stochastic processes, e.g. filtration and stopping time \cite{probzoo}. 
Similarly, the HOL4 library contains a general framework for probability theory including 
the formalization of the laws of large numbers along side results such as 
the Borel-Cantelli lemmas and Kolmogorov's 0-1 law \cite{hol4description}. Parts of probability theory 
has also been formalized in Mizar \cite{mizar} and Coq \cite{affeldt2012itp, polariscoq} focusing on 
discrete/finite sample spaces. Probability theory in Lean as with other theories is consolidated in \lstinline{mathlib}. 
Built upon the measure theory part of the library, \lstinline{mathlib} contains probability results ranging from 
the strong law of large numbers to parts of the portmanteau theorem alongside theorems introduced through this 
project (see the \lstinline{mathlib} overview page\footnote{\href{https://leanprover-community.github.io/mathlib-overview.html}{https://leanprover-community.github.io/mathlib-overview.html}} for more details).

\subsection{Martingales and the convergence theorems}
\label{sub:martingales_and_the_convergence_theorems}

\paragraph{Measurable spaces and measures}

This work concerns objects defined in a measure space $(\Omega, \mathscr A, \mu)$, that is a type (or set) $\Omega$, a $\sigma$-algebra $\mathscr A$ and a measure on that $\sigma$-algebra $\mu$. A $\sigma$-algebra is a collection of subsets of $\Omega$ closed under complement, countable unions, and countable intersections. The sets in $\mathscr A$ are called measurable sets.
A measure is a map from $\mathscr A$ to $\mathbb{R}_{\ge 0}\cup \{\infty\}$ with value zero on the empty set and which verifies a countable additivity assumption. For a measure $\mu$, there is a standard way to give a value $\mu(s)$ to any subset $s$ of $\Omega$, not only the sets in $\mathscr A$ (through outer measures, which we will not define). 

We will say that a property $p:\Omega\to Prop$ is true almost everywhere (a.e.) if $\mu (\{\omega \mid \neg p(\omega)\}) = 0$, and denote this in \lstinline{mathlib} by \lstinline{∀ᵐ ω ∂μ, p ω}. Similarly, we define almost everywhere equality of functions (or inequalities): \lstinline{f =ᵐ[μ] g} means that $f$ and $g$ are almost everywhere equal.

A measure $\mu$ is said to be finite if $\Omega$ has finite measure: $\mu(\Omega) < \infty$. It is $\sigma$-finite if there is an increasing sequence of measurable sets $(S_n)_{n \in \mathbb{N}}$ with finite measure such that $\bigcup_{n \in \mathbb{N}} S_n = \Omega$. If $\mu(\Omega) = 1$, then $\mu$ is said to be a probability measure.
This paper is about objects like stochastic processes, which are most often used in the context of probability theory. However none of our results will require that $\mu$ is a probability measure.
We will try to state results in the greatest possible generality. Following that principle, many lemmas will not use any assumption on the measure, while other will require it to be $\sigma$-finite, or even finite. We never use probability measures because such a strong assumption is nowhere needed.
This approach is very different from the one of \cite{vajjha2022formalization} for example, in which one of the design choice of the probability library is to only consider probability measures. 

\paragraph{Random variables and spaces of functions}

A random variable $f:\Omega \to E$ is a function from $\Omega$ to another space $E$. In most places, we will suppose that $E$ is a Banach space (complete real normed vector space), because that assumption allows us to define a notion of integral called the Bochner integral, which will be the main theory of integration we use in this paper.

\lstinline|mathlib| contains several notions of measurability, and we now explain the ones which are most relevant to our work: \lstinline|strongy_measurable| and \lstinline|ae_strongly_measurable|. A simple function is defined as a piecewise constant function taking finitely many values, such that the preimage of each value is a measurable set. We denote the set of such functions by $\mathcal S(E, \mathscr A)$. A function is said to be strongly measurable if it is the limit of a sequence of simple functions. It is said to be a.e. strongly measurable if it is a.e. equal to a strongly measurable function.

In second-countable metrizable spaces with the Borel $\sigma$-algebra, ``strongly measurable'' and the more commonly used ``measurable'' are equivalent, hence a reader accustomed to that second notion who would be mostly interested in applications to spaces like $\mathbb{R}^d$ can safely ignore the difference.

We denote by $L^0(E, \mu)$ the space of a.e. strongly measurable functions, quotiented by the a.e. equal relation. The objects of that space are rigorously speaking classes of functions, but we still use the word function.
For $p \ge 1$, we denote by $L^p(E, \mu)$ the subset of $L^0(E, \mu)$ containing the functions whose norm raised to the power $p$ has finite Lebesgue integral (denoted by $\int^-$), that is
\begin{align*}
L^p(E, \mu) = \left\{f \in L^0(E, \mu) \mid \int_x^- \lVert f(x) \rVert^p d \mu(x) < \infty\right\} \: .
\end{align*}
The functions in $L^1(E, \mu)$ are said to be integrable. On this $L^1$ space the Bochner integral is well defined and we can talk about the integral of $f$, $\int_x f(x)d\mu(x)$, for which we also introduce the notation $\mu[f]$ in Lean. When $\mu$ is a probability measure, this integral is also called expectation of $f$.

\paragraph{Stochastic processes, filtrations and martingales}

A stochastic process $(f_i)_{i \in \mathcal \iota}$ is an \(\iota\)-indexed family of random variables, where the index set is ordered and usually represents time. Most often $\iota = \mathbb{N}$ or $\iota = \mathbb{R}_{\ge 0}$.

Let us consider a simple example of a stochastic process: the amount of money of a player who bets on coin flips. At time $t \in \mathbb{N}$, a coin is flipped and the player gains 1 if the result of the flip is heads, and loses 1 if it is tails. The amount of money of the player over time is a stochastic process $(f_n)_{n\in \mathbb{N}}$, where each $f_n$ is a real random variable $\Omega \to \mathbb{R}$.

A filtration $(\mathscr{F}_i)_{i\in \iota}$ is a non-decreasing sequence of sub-$\sigma$-algebras, and often represents information available up to some time. In our coin flip example, let $X_i$ be the random variable describing the coin flip at time $i$, and let $\sigma(X_i)$ be the $\sigma$-algebra generated by $X_i$. We can consider the filtration defined by $\mathscr{F}_i = \mathscr{F}_{i-1} \vee \sigma(X_i)$ (i.e. the smallest \(\sigma\)-algebra generated by \(\mathscr{F}_{i-1}\) and \(X_i\)) with $\mathscr{F}_0$ the trivial $\sigma$-algebra, which encodes the observations available after flip $i$.
A stochastic process is said to be adapted to a filtration if for all $i \in \iota$, $f_i$ is $\mathscr{F}_i$-strongly measurable (strongly measurable with respect to the $\sigma$-algebra $\mathscr{F}_i$). Intuitively, the capital of the player after $i$ coin flips can be computed from the knowledge of all the results up to time $i$.

While the expectation $\mu[f_n]$ for a probability measure $\mu$ can be interpreted as the expected value of $f_n$, which equals the initial capital $f_0$ in our example for fair coins since we expect heads and tails to happen equally often; the conditional expectation $\mu[f_n \mid \mathscr{F}_i]$ is the value of $f_n$ we expect once we have seen the $i$ first coin tosses.
If the coins are fair, the process describing the amount of money of the player verifies $\mu[f_n \mid \mathscr{F}_{i}] = f_{i}$ almost everywhere for all $i \le n$: such a process is called a martingale.

\paragraph{The martingale convergence theorems}

The two martingale convergence theorems we formalize give conditions under which a (sub)martingale $(f_n)_{n \in \mathbb{N}}$ has a limit function $f$ as $n$ tends to infinity. The first one gives almost everywhere convergence to that limit. 

\begin{theorem*}[Almost everywhere martingale convergence]
  Given a submartingale \((f_n)_{n \in \mathbb{N}}\) adapted to the filtration \((\mathscr{F}_n)_{n \in \mathbb{N}}\), 
  if there exists some \(R \in \mathbb{R}\) such that \(\|f_n\|_1 := \int |f_n| \le R\) for all \(n \in \mathbb{N}\), then there 
  exists some \(\bigvee_{n \in \mathbb{N}} \mathscr{F}_n\)-measurable function \(f\) which is integrable such that 
  \(f_n \xrightarrow{\text{a.e.}} f\).

  We will for shorthand denote \(\bigvee_{n \in \mathbb{N}} \mathscr{F}_n\) by \(\mathscr{F}_\infty\).
\end{theorem*}

The second result we prove characterizes the $L^1$ convergence of martingales.
\begin{theorem*}[\(L^1\) martingale convergence]
  Given \((f_n)_{n \in \mathbb{N}}\) a martingale adapted to the filtration \((\mathscr{F}_n)_{n \in \mathbb{N}}\), 
  the following are equivalent:
  \begin{enumerate}[(i)]
    \item \((f_n)\) is uniformly integrable (in the probability sense);
    \item \((f_n)\) converges in \(L^1\) to some \(\mathscr{F}_\infty\)-measurable, integrable function \(f\);
    \item there exists some \(\mathscr{F}_\infty\)-measurable and integrable \(f\) such that for all \(n \in \mathbb{N}\), 
    $\mu[f \mid \mathscr{F}_n] = f_n \text{ a.e.} \: .$
  \end{enumerate}
\end{theorem*}

Those two theorems are at the core of the study of martingales, which are themselves very versatile tools in probability theory. They can for example be used as in \cite{williams1991probability} to prove Kolmogorov's 0-1 law, the strong law of large numbers or the generalized Borel-Cantelli lemma.

\subsection{Lean and mathlib}
\label{sub:lean_and_mathlib}

All the code produced for this paper was contributed to \lstinline{mathlib} \cite{mathlib}: the mathematical library for the Lean theorem prover \cite{moura2015lean}. This library contains formalized definitions and theorems from many domains of mathematics, in a way that ensures that all those definitions are usable together without compatibility issues. We will see in the section on the definition of the conditional expectation that this feature was essential to our work. Probability theory is built on a large corpus of results in analysis, topology, descriptive set theory and measure theory, and its formalization beyond discrete spaces requires a library which is well developed in all those areas.

New code submitted to \lstinline{mathlib} goes through a review process supervised by the \lstinline{mathlib} maintainers, helped by automatic checks (for example checking for unused arguments in theorems) as described in \cite{doorn2020maintaining}. Ensuring that the new code is compatible with and general enough to be used by other parts of the library is a central part of that review process.

Our paper can be understood without prior knowledge of Lean syntax, especially since it contains only Lean statements and no proofs. These statements however contain three distinct types of parameters, which we now explain on an example.
In the remainder of the paper we assume that a measure space $(\Omega, \mathscr A, \mu)$ is given, as well as an index set $\iota$ and a Banach space $E$ for the values of the random processes. To obtain the same effect as that sentence in Lean, we would add the following line at the top of the file:
\begin{lstlisting}
variables {Ω ι E : Type*} {𝒜 : measurable_space Ω}
  {μ : measure Ω} [normed_add_comm_group E]
  [normed_space ℝ E] [complete_space E]
\end{lstlisting}
Then those parameters don't need to be given to every lemma and definition, and we will also not add them to the Lean code we write below. The parameters enclosed in curly brackets like \lstinline|{μ : measure Ω}| are implicit. Lean is supposed to infer them from other hypotheses: the measure will appear in other hypotheses of the theorems and Lean should guess it from there.
An explicit parameter would be given in parentheses, for example \lstinline|(μ : measure Ω)|, and the user would have to provide it when they want to apply the theorem. In the above variables definition, we also see \lstinline|[complete_space E]|, which is an instance parameter: Lean uses typeclass resolution to deduce those parameters from the definition of $E$ and other instances in the hypotheses. If for example $E = \mathbb{R}$, we won't need to supply a proof that $\mathbb{R}$ is complete, since typeclass resolution can build it automatically. For a detailed discussion of instance parameters in \lstinline{mathlib}, see \cite{baanen2022use}.

\section{Stopping times and martingales: definitions}
\label{sec:stopping_times_and_martingales_definitions}

In order to prove the martingale convergence theorems, we first need to formalize the objects they use, namely conditional expectations, uniform integrability, stochastic processes, stopping times and martingales. This section is dedicated to our development of these basic objects of probability theory.

\subsection{Conditional expectation}
\label{sub:conditional_expectation}
% !TeX root = ../martingales_in_lean.tex

The definition of martingales requires a notion of conditional expectation, which is a fundamental tool in probability.

\begin{definition}\label{def:condexp}
Let $(\Omega, \mathscr A, \mu)$ be a measure space and let $\mathscr B$ be a sub-$\sigma$-algebra of $\mathscr A$. Let $E$ be a Banach space (complete normed vector space) and $f : \Omega \to E$ be an integrable function (with respect to $\mu$). Then a function $g : \Omega \to E$ is said to be a \emph{conditional expectation} of $f$ with respect to $\mathscr B$ if it is integrable, $\mathscr B$-strongly measurable and for all $\mathscr B$-measurable sets $s$,
\begin{align*}
\int_{x \in s} g(x) d\mu(x) = \int_{x \in s} f(x) d\mu(x) \: .
\end{align*}
\end{definition}

If the measure $\mu$ restricted to the $\sigma$-algebra $\mathscr B$ is $\sigma$-finite, then there exists a conditional expectation and it is unique up to almost everywhere equality.

Our formalization broadly follows the procedure described in \cite{hytonen2016analysis} and leads to a definition with the following type.
\begin{lstlisting}
def condexp (ℬ : measurable_space Ω)
  {𝒜 : measurable_space Ω}
  (μ : measure Ω) (f : Ω → E) : Ω → E :=
if hm : ℬ ≤ 𝒜
  then if h : sigma_finite (μ.trim hm) ∧ integrable f μ
    then if strongly_measurable[ℬ] f then f
      else (@ae_strongly_measurable'_condexp_L1 _ _ _ _ _ ℬ 𝒜 μ hm h.1 _).mk (@condexp_L1 _ _ _ _ _ _ _ hm μ h.1 f)
    else 0
  else 0
\end{lstlisting}

We will always try to make the use of the library as close as possible to paper proofs, and to that purpose we define a notation \lstinline{μ[f | ℬ]} for \lstinline{condexp ℬ μ f}.

\begin{lstlisting}
notation μ `[` f `|` ℬ `]` := measure_theory.condexp ℬ μ f
\end{lstlisting}

Let us parse the definition. $\Omega$ is a measurable space with $\sigma$-algebra $\mathscr{A}$ on which we have a measure $\mu$ and $E$ is a Banach space. Then $\mathscr B$ is the $\sigma$-algebra with respect to which we are computing the conditional expectation and $f$ is the function we are computing it for. What we obtain is a function $\Omega \to E$.

In contrast to definition~\ref{def:condexp}, the Lean definition does not require $\mathscr B$ to be a sub-$\sigma$-algebra of $\mathscr{A}$, nor does it require $f$ to be integrable or the measure to be $\sigma$-finite on $\mathscr B$. Instead, we define $\mu[f \mid \mathscr B]$ as 0 if those conditions are not verified.
Our function \lstinline{condexp} is then a true conditional expectation only under additional assumptions, but defining it this way allows us to write $\mu[f \mid \mathscr B]$ without having to add to that notation proofs of the three required properties.
Defining such global functions with default values is a common strategy in theorem provers to make the definitions easier to use. Of course the hypotheses we omitted are then required for most lemmas describing the properties of \lstinline{condexp}.

The \lstinline{condexp_L1} auxiliary function is the result of the construction detailed in the next section. We also check whether $f$ is $\mathscr B$-strongly measurable: in that case, any $\mathscr B$-strongly measurable function a.e. equal to $f$ is a conditional expectation, and \lstinline{condexp_L1} will be one of those, but we choose to enforce everywhere equality instead of a.e. equality to ease the use of the function.
Indeed, while the properties of the conditional expectation we will be interested in are invariant to a.e. equality and thus the choice of a particular function changes nothing mathematically, having exact equality to $f$ in the strongly measurable case makes it easier to replace $\mu[f | \mathscr B]$ by $f$ in the code.

Among the many properties of \lstinline{condexp}, we most notably prove the equality of integrals which characterizes conditional expectations:
\begin{lstlisting}
theorem set_integral_condexp (hB : ℬ ≤ 𝒜)
  [sigma_finite (μ.trim hB)] (hf : integrable f μ)
  (hs : measurable_set[ℬ] s) :
∫ x in s, μ[f | ℬ] x ∂μ = ∫ x in s, f x ∂μ
\end{lstlisting}
This proves that \lstinline{μ[f | ℬ]} is indeed the conditional expectation of $f$.
The generality in which we define the conditional expectation is close to the greatest possible one: the only condition which is more relaxed in \cite{hytonen2016analysis} compared to our implementation is the integrability.
We require the function \lstinline|f| to be integrable to have a non-zero \lstinline{μ[f | ℬ]}, while \cite{hytonen2016analysis} define a meaningful conditional expectation for the larger class of $\sigma$-integrable functions, which means that there is an increasing sequence of sets with union equal to the whole space such that the function is integrable on each individual set.
$\sigma$-integrability is not defined in \lstinline|mathlib| and we chose not to add it since it is rarely used in the literature.

Defining the conditional expectation for Banach space and $\sigma$-finite measure can be useful in other fields than probability, for example in harmonic analysis where the measure is often the Lebesgue measure, which is not finite.

\paragraph{Construction of the conditional expectation}

The main goal of our construction is to obtain the conditional expectation as a continuous linear map from $L^1(E, \mu)$ to itself.
The final definition is then obtained from that map by wrapping it in a few if...then...else checks as explained above.

The first step in that construction is to define a conditional expectation for indicators of measurable sets with finite measure.
An indicator of a measurable set with finite measure is a function in $L^2(E, \mu)$, so we can focus on those. We want the conditional expectation to be a $\mathscr B$-strongly measurable function, and those functions form a linear subspace of $L^2(E, \mu)$.
We can then obtain a conditional expectation for any function in $L^2(E, \mu)$ by projecting orthogonally on the subspace of $\mathscr B$-strongly measurable functions.
The existence of such an orthogonal projection, the linearity properties and the fact that the projection verifies the conditional expectation characterization are all consequences of properties of projections in inner product spaces which were already available in the analysis part of \lstinline|mathlib| and directly applicable. The nature of \lstinline|mathlib| as a monolithic library containing all kinds of mathematical domains made the following quick definition possible:
\begin{lstlisting}
def condexp_L2 (hm : ℬ ≤ 𝒜) :
  (Ω →₂[μ] E) →L[ℝ] (Lp_meas E ℝ ℬ 2 μ) :=
@orthogonal_projection ℝ (Ω →₂[μ] E) _ _
  (Lp_meas E ℝ ℬ 2 μ)
  (by { haveI : fact (ℬ ≤ 𝒜) := ⟨hm⟩, apply_instance, })
\end{lstlisting}
That \lstinline|condexp_L2| is a linear map from $L^2(E, \mu)$ (denoted by \lstinline|Ω →₂[μ] E|) to the type \lstinline|Lp_meas E ℝ ℬ 2 μ|, which is the subspace of $L^2(E, \mu)$ which contains a.e. $\mathscr B$-strongly measurable functions. It is defined by using \lstinline|orthogonal_projection|. The last argument is a proof that the subspace is complete: that proof first makes the fact that $\mathscr B$ is a sub-$\sigma$-algebra of $\mathscr{A}$ available to the type class resolution mechanism, then calls the \lstinline|apply_instance| tactic, which uses that mechanism to generate a proof.

We now explain how we can extend that definition which works in particular for indicators of measurable sets with finite measure, to obtain a linear map from $L^1(E, \mu)$ to itself, which will give us the full conditional expectation.

Let $E$ be a normed real vector space and $G$ be a Banach space. Let $(\Omega, \mathscr A, \mu)$ be a measure space. Suppose that we have a map $T$ from the subsets of $\Omega$ to the continuous linear maps from $E$ to $G$ which verifies the following properties:
\begin{enumerate}
	\item for any two disjoint measurable sets with finite measure $s,t$, $T (s \cup t) = T(s) + T(t)$,
	\item there exists $C \ge 0$ such that for any measurable set $s$ with finite measure, $\lVert T(s) \rVert \le C \mu(s)$.
\end{enumerate}
Then we can define a continuous linear map $T^{L^1}$ from $L^1(E, \mu)$ to $G$, such that for all measurable sets with finite measure $s$ and all $x \in E$, $T^{L^1}$ applied to the indicator of the set $s$ with value $x$ is equal to $T(s,x)$.

The extension process is done in several successive phases.
First, the function can naturally be defined on indicators of measurable sets with finite measure by $T(s,x)$.
Then thanks to the linearity property of $T$ it can be extended to integrable simple functions in $\mathcal S(E, \mathscr A)$, which are linear combinations of such indicators.
It can then be extended to functions which are almost everywhere equal to an integrable simple function $f_s$ by taking the value of $T$ on $f_s$.
The second property of $T$ (dominated norm) is used to prove that the resulting map from a.e. classes of integrable simple functions to $G$ is a continuous linear map.
Since that map is continuous, we can give a unique value to functions which are a.e. limits of sequences of integrable simple functions. Those are the a.e. strongly measurable and integrable functions, e.g. functions in $L^1(E, \mu)$, to which we can then finally extend our map.

This extension process is added to \lstinline|mathlib| and is used for two applications:
\begin{itemize}
	\item For $T(s,x)$ the conditional expectation of the indicator of set $s$ with value $x$, with value in $G = L^1(E, \mu)$, we get for $T^{L^1}$ the conditional expectation of a function in $L^1(E, \mu)$, as a linear map from $L^1(E,\mu)$ to $L^1(E,\mu)$.
	\item For $T(s,x) = \mu(s) \cdot x$ with value in $G = E$, $T^{L^1}$ maps a function in $L^1(E, \mu)$ to its Bochner integral.
\end{itemize}
A special case of the extension process was previously implemented for the definition of the Bochner integral, but we refactored the library to isolate this general construction and apply it to both the definition of the integral and the conditional expectation.

With this refactor we were able to reuse many proofs first written for integrals, up to appropriate generalization, and obtain relatively quickly the corresponding properties for \lstinline|condexp|. Like many parts of \lstinline|mathlib|, the integration files were subject to frequent changes in the last year and the integral was slightly altered or generalized several times to allow an easier use in other parts of the library. See \cite{gouezel} for an overview of the other recent integral refactors.

\subsection{Uniform integrability}
\label{sub:unif_int}
% !TeX root = ../martingales_in_lean.tex

We now define uniform integrability, which will feature in the $L^1$ convergence theorem.
In the literature, there are two non-equivalent formulations of uniform integrability used 
by analysts and probabilists respectively. 

\begin{definition}[Uniform integrability, analyst's definition]
  A sequence of functions \((f_n) \subseteq L^1(E,\mu)\) is said to be uniformly integrable 
  if for all \(\epsilon > 0\), there exists some \(\delta > 0\) such that for all 
  \(A \in \mathscr{A}\) with \(\mu(A) < \delta\), we have for all \(n\),
  \[\|f_n \mathbf{1}_A\|_1 < \epsilon.\]
  Where we recall \(\|f\|_1 = \int_x \|f(x)\| d \mu(x)\). 
  
\end{definition}

More generally, for $p \ge 1$ we define the norm 
$\lVert f \rVert_p = (\int_x \lVert f (x)\rVert^p d \mu(x))^{1/p}$ on \(L^p(E, \mu)\). 
We say that a sequence of \(L^p(E, \mu)\) functions is said to converge in $L^p$ if it converges with 
respect to the topology induced by this norm.

\begin{definition}[Uniform integrability, probabilist's definition]
  A sequence of functions \((f_n) \subseteq L^1(E,\mu)\) is said to be uniformly integrable if
  \[\lim_{C \to \infty} \sup_n \|f_n \mathbf{1}_{\{|f_n| \ge C\}}\|_1 = 0.\] 
\end{definition}

Uniform integrability establishes a relation between convergence in measure and convergence 
in \(L^p\) known as the Vitali convergence theorem. This is the main result required to prove the 
\(L^1\) martingale convergence theorem.

\begin{theorem}[Vitali convergence theorem]
  A sequence of functions \((f_n) \subseteq L^p(E, \mu)\) converges in \(L^p\) if and only if 
  \((f_n)\) converges in measure and \((|f_n|^p)\) is uniformly integrable in the analyst's sense.
\end{theorem}

The Vitali convergence theorem also holds for uniform integrability in the probabilist's sense as 
the probabilist's definition is strictly stronger than the analyst's definition. This is made clear 
by the following proposition.

\begin{proposition}\label{prop:unif_int_prob}
  If \(\mu\) is a finite measure, then \((f_n) \subseteq L^1(E,\mu)\) is uniformly integrable in the 
  probabilist's sense if and only if it is uniformly bounded in \(L^1\) and is uniformly integrable 
  in the analyst's sense.
\end{proposition}

Hence, it is equivalent to define probabilist's uniform integrability with the characterization of
proposition~\ref{prop:unif_int_prob}. This alternative characterization has the advantage of 
directly relating the two definitions and so, allowing us to only make one set of API without 
needing too many translation lemmas. 

In Lean, these two definitions are formulated as,
\begin{lstlisting}
def unif_integrable {𝒜 : measurable_space α} 
  (f : ι → α → β) (p : ℝ≥0∞) (μ : measure α) :=
∀ (ε : ℝ) (hε : 0 < ε), ∃ (δ : ℝ) (hδ : 0 < δ), 
  ∀ i s, measurable_set s → μ s ≤ ennreal.of_real δ →
  snorm (s.indicator (f i)) p μ ≤ ennreal.of_real ε

def uniform_integrable {𝒜 : measurable_space α}
  (f : ι → α → β) (p : ℝ≥0∞) (μ : measure α) :=
(∀ i, ae_strongly_measurable (f i)) 
  ∧ unif_integrable f p μ 
  ∧ ∃ C : ℝ≥0, ∀ i, snorm (f i) p μ ≤ C
\end{lstlisting}
for the analyst's and probabilist's definition respectively where \lstinline|snorm f p μ| is the 
function evaluating to \(\|f\|_p\) for some suitable \(f\) with respect to the measure \(\mu\).

We note that the Lean formulation of both definitions has an extra parameter 
\lstinline{p : ℝ≥0∞} for which our original definition is recovered if \(p = 1\). 
This allows us to talk about uniform integrability in the case \((f_n)\) is not necessarily 
\(L^1\). In the case that the measure \(\mu\) is finite, by Hölder's inequality, we have 
\begin{lstlisting}
p ≤ q → 
  uniform_integrable f q μ → uniform_integrable f p μ
\end{lstlisting}

To recover the original definition of uniform integrability in the probabilist's sense, we 
proved the following:
\begin{lstlisting}
lemma uniform_integrable_iff 
  [is_finite_measure μ] (hp : 1 ≤ p) (hp' : p ≠ ∞) :
uniform_integrable f p μ 
  ↔ (∀ i, ae_strongly_measurable (f i)) 
  ∧ ∀ ε : ℝ, 0 < ε → ∃ C : ℝ≥0,
  ∀ i, snorm ({x | C ≤ ∥f i x∥₊}.indicator (f i)) p μ 
    ≤ ennreal.of_real ε
\end{lstlisting}
where \lstinline|∥x∥₊| denotes the non-negative real valued norm of any suitable \(x\).

As we included the extra parameter \lstinline{p} in our formalization of uniform 
integrability, in contrast to the formulation of the Vitali convergence theorem above, we do not need to 
say ``\((|f_n|^p)\) is uniformly integrable'' but simply 
``\((f_n)\) is \(p\)-uniformly integrable'' resulting in the following formalization of the Vitali convergence theorem,
\begin{lstlisting}
lemma tendsto_in_measure_iff_tendsto_Lp 
  [is_finite_measure μ] (hp : 1 ≤ p) (hp' : p ≠ ∞) 
  (hf : ∀ n, mem_ℒp (f n) p μ) (hg : mem_ℒp g p μ) :
tendsto_in_measure μ f at_top g 
  ∧ unif_integrable f p μ
  ↔ tendsto (λ n, snorm (f n - g) p μ) at_top (𝓝 0)
\end{lstlisting}
with \lstinline{tendsto_in_measure μ f at_top g} being the formalization of  
``\((f_n)\) converges in measure along \(\mu\) to \(g\) as \(n \to \infty\)'',
\lstinline{tendsto (λ n, snorm (f n - g) p μ) at_top (𝓝 0)} the formalization of 
``\(\|f_n - g\|_p\) converges to 0 as \(n \to \infty\)'', and 
\lstinline{mem_ℒp g p μ} the assertion that \(g\) is a.e.-measurable and \(\|g\|_p < \infty\).

Our first implementation of uniform integrability in the probability sense required that the functions are 
strongly measurable instead of a.e. strongly measurable.
The initial consensus among interested \lstinline|mathlib| community members was that the probability part of the library could focus 
solely on functions which are strongly measurable, and did not need to use the a.e. strongly measurable notion.
However, later work on the \(L^p\)-version of the strong law of large numbers (SLLN)
indicated that this definition of uniform integrable is too strong. 

The a.e.-version of the SLLN states that, given a sequence of independent and identically 
distributed (i.i.d.) random variables \((f_n)_{n \in \mathbb{N}}\) with \(\|f_0\|_1 < \infty\), 
\(S_n := \frac{1}{n} \sum_{i = 1}^n f_n\) converges to \(\mu[f_0]\) almost everywhere. Hence, as a sequence of identically distributed 
random variables is uniformly integrable, and as the averaging of a uniformly integrable 
sequence is also uniformly integrable, the \(L^p\)-version of the SLLN follows by 
the Vitali convergence theorem, i.e. for i.i.d. random variables \((f_n)_{n \in \mathbb{N}}\) with 
\(\|f_0\|_p < \infty\), \(S_n\) converges to \(\mu[f_0]\) in \(L^p\).

In the outlined proof above, we use the following lemma.
\begin{lstlisting}
lemma uniform_integrable_of_ident_distrib {f : ι → α → E}
  {j : ι} {p : ℝ≥0∞} (hp : 1 ≤ p) (hp' : p ≠ ∞)
  (hℒp : mem_ℒp (f j) p μ) 
  (hf : ∀ i, ident_distrib (f i) (f j) μ μ) :
  uniform_integrable f p μ
\end{lstlisting}
We note that, in the case uniform integrability only requires a.e. strongly measurability, the 
above assumptions are sufficient. Indeed, as we have asserted that \(f_j\) is $L^p$, \(f_j\) is a.e. strongly 
measurable by the definition of \lstinline{mem_ℒp}. Hence, by the properties of identically distributed 
functions, \(f_i\) is also a.e. strongly measurable for all \(i \in \iota\). In contrast, should uniform 
integrability require strongly measurable, as a function identically distributed with a strongly measurable functions 
is not automatically strongly measurable, we will in addition require \(f_i\) to be strongly measurable for all \(i\) 
rather than only imposing measurability of \(f_j\). 

In essence, we are manipulating properties of \(L^p\) classes and any predicate \(P\) on them should apply 
to all elements of said class; i.e. \(P\) should factor through the following triangle: 
\[\begin{tikzcd}
	{\mathcal{L}^p:=\left\{f \text{ a.e. measuable} \mid \|f\|_p < \infty\right\}} & {\text{Prop}} \\
	\\
	{L^p:=\mathcal{L}^p / \text{a.e. equality}}
	\arrow["P", from=1-1, to=1-2]
	\arrow["Q"', from=1-1, to=3-1]
	\arrow["{\tilde P}"', from=3-1, to=1-2]
\end{tikzcd}\]
with \(Q\) being the quotient map. With this in mind, we will allow a.e. strongly measurable functions 
in our discourse.

\subsection{Stochastic processes}
\label{sub:stopping}
% !TeX root = ../martingales_in_lean.tex

A stochastic process is a sequence of random variables evolving in time. 
While this project mostly deals with discrete time processes, we will in general not to be too 
restrictive on the time indices while making definitions, e.g. we will allow the time index to be an 
arbitrary set equipped with some ordering in most cases.  

Given a measurable space \((\Omega, \mathscr{A})\), a filtration on this measurable space is a 
sequence of increasing sub-\(\sigma\)-algebras of \(\mathscr{A}\): \((\mathscr{F}_i)_{i \in \iota}\) with 
some time index set \(\iota\) (most commonly \(\mathbb{N}\) or \(\mathbb{R}_{\ge 0}\)). 
\begin{lstlisting}
structure filtration {Ω : Type*} (ι : Type*) [preorder ι] 
  (𝒜 : measurable_space Ω) :=
(seq : ι → measurable_space Ω)
(mono' : monotone seq)
(le' : ∀ i : ι, seq i ≤ 𝒜)
\end{lstlisting}
In Lean, we define filtrations as a structure and prove that it forms a complete lattice with respect
to the order inherited from the order on \(\sigma\)-algebras.
\begin{lstlisting}
instance : complete_lattice (filtration ι 𝒜)
\end{lstlisting}
With this, given a filtration \((\mathscr{F}_i)_{i \in \iota}\) and a stochastic process 
\((f_i)_{i \in \iota}\), we say \((f_i)\) is adapted to \((\mathscr{F}_i)\) if \(f_i\) is 
\(\mathscr{F}_i\)-strongly measurable for all \(i \in I\). 
\begin{lstlisting}
def adapted (ℱ : filtration ι 𝒜) (f : ι → Ω → β) : Prop :=
∀ i : ι, strongly_measurable[ℱ i] (f i)
\end{lstlisting}

In stochastic processes, it is useful to be able to talk about not just deterministic times but 
also random times as well. To achieve this we introduce the notion of stopping times. Stopping times 
are a useful definition to characterize the times at which a process exhibits some given special property. 
To prove the martingale convergence theorems, we develop a theory of stopping times in Lean and
define an important class of stopping time known as hitting time\footnote{While hitting times are not in general stopping times, 
they are in the case that the time index is discrete which is the setting we are working under.}.

Given some \(\iota\)-indexed filtration \((\mathscr{F}_i)_{i \in \iota}\), we say a function 
\(\tau : \Omega \to \iota\) is a stopping time if for all \(i \in \iota\), 
\(\{\tau \le i\}\) is \(\mathscr{F}_i\)-measurable. 
\begin{lstlisting}
def is_stopping_time [preorder ι] (ℱ : filtration ι 𝒜) 
  (τ : Ω → ι) :=
∀ i : ι, measurable_set[ℱ i] {ω | τ ω ≤ i}
\end{lstlisting}

Given some \(E\)-valued process \((f_n)_{n \in \mathbb{N}}\) and a set \(A \subseteq E\), the 
hitting time is thought of as the first time \((f_n)\) enters \(A\). Namely, it is defined 
to be 
\[H^f(A) : \Omega \to \mathbb{N} \cup \{\infty\} : 
  \omega \mapsto \inf \{n \mid f_n(\omega) \in A\}\]
where we define \(\inf \emptyset = \infty\). Given this definition, it is important in our proof of 
the martingale convergence theorems that whenever \(\{n \mid f_n(\omega) \in A\} \neq \emptyset\) we observe
\begin{enumerate}
  \item \(f_{H^f(A)} \in A\) and,
  \item \(\forall n < H^f(A), f_n \not\in A\).
\end{enumerate}
In particular, taking the stopped process 
defined by \(N \in \mathbb{N} \mapsto f_{H^f(A) \wedge N}\) where \(N \wedge \infty = N\),
we have \(f_{H^f(A)} \in A\) and for all \(n < H^f(A)\), \(f_n \not\in A\) if \(H^f(A) < N\). 
This is however not possible in a typed system such as Lean since \(f_{H^f(A)}\) does not 
type check as \((f_n)\) is indexed by \(\mathbb{N}\) while \(H^f(A)\) has type  
\(\mathbb{N} \cup \{\infty\}\) (in contrast to on paper where we prove the definition is well-defined 
after the fact by showing \(N \wedge \infty \in \mathbb{N}\) for all \(N \in \mathbb{N}\)). 
Thus, an alternative formulation of the hitting time is required. 

A naïve approach is to directly take the range of \(H^f(A)\) to be \(\mathbb{N}\) where 
\(\inf \emptyset\) is defined to be some natural number \(n\) (in Lean \(n\) is taken to be \(0\)). 
This approach however, does not allow us to reobtain the two properties required for the martingale 
convergence theorem since for \(N > n\), \(\inf \emptyset \wedge N = n < N\) and hence, is not 
suitable for our purpose. 

Instead, we would like to incorporate the stopped process notion into our definition of the hitting 
time. Namely, by introducing an ending time \(N\), we define the hitting time to be the 
first time \((f_n)\) enters \(A\) before time \(N\). Furthermore, in order for our definition to 
also work in the case that the time index is unbounded from below, it is also useful to introduce 
a starting time, resulting in the following definition
\begin{lstlisting}
def hitting [preorder ι] [has_Inf ι] 
  (f : ι → α → β) (s : set β) (n m : ι) : α → ι :=
λ x, if ∃ j ∈ set.Icc n m, f j x ∈ s then 
  Inf (set.Icc n m ∩ {i : ι | f i x ∈ s}) else m
\end{lstlisting}
where \lstinline{set.Icc n m} denotes the closed interval \([n, m]\). 
We introduce th notation \(H_{N, M}^f(A)\) for the hitting time of \(f\) in \(A\) with starting time 
\(N\) and ending time \(M\).

The lemma that discrete hitting times are stopping times is formalized as 
\begin{lstlisting}
lemma hitting_is_stopping_time
  [topological_space β] [pseudo_metrizable_space β] 
  [measurable_space β] [borel_space β]
  {f : filtration ℕ 𝒜} {u : ℕ → Ω → β} {s : set β} {n n' : ℕ}
  (hu : adapted f u) (hs : measurable_set s) :
is_stopping_time f (hitting u s n n')
\end{lstlisting}

\subsection{Martingales}
\label{sub:mgale}
% !TeX root = ../martingales_in_lean.tex

\begin{definition}[\cite{hytonen2016analysis}, definition 3.1.1]\label{def:martingale}
A family of integrable functions $(f_i)_{i \in \iota}$ with values in a Banach space $E$ is called a martingale with respect to a $\sigma$-finite filtration $(\mathscr F_i)_{i\in \iota}$ if it is adapted to $(\mathscr F_i)$ and for all $i, j \in \iota$ with $i \le j$,
$\mu[f_j \mid \mathscr F_i] = f_i$ almost everywhere.
\end{definition}

If there is an order $\le$ defined on $E$, then we can also define \emph{submartingales} and \emph{supermartingales}, with the same definition except that the equality of the conditional expectation is replaced respectively by $\mu[f_j \mid \mathscr F_i] \ge f_i$ and $\mu[f_j \mid \mathscr F_i] \le f_i$.
Martingales are both submartingales and supermartingales.
Our Lean definition of a martingale is
\begin{lstlisting}
def martingale [preorder ι] (f : ι → Ω → E)
  (ℱ : filtration ι 𝒜) (μ : measure Ω) : Prop :=
adapted ℱ f ∧ ∀ i j, i ≤ j → μ[f j | ℱ i] =ᵐ[μ] f i
\end{lstlisting}
\lstinline{submartingale} and \lstinline{supermartingale} are defined similarly, under the additional hypothesis \lstinline{[has_le E]} which requires that $E$ has an order.

This definition inherits the generality of the conditional expectation construction: $E$ is any Banach space. The index set can also be any preorder, which covers in particular the two usual cases of discrete and continuous time indexes ($\mathbb{N}$ and $\mathbb{R}$).
The function sequence $(f_i)$ is not required to contain integrable functions (in contrast to Definition~\ref{def:martingale}), but if $f_j$ is not integrable then $\mu[f_j | \mathscr F_i] = 0$ (by definition) and the sequence has to be zero for all $i \le j$.
Similarly to the definition of the conditional expectation, the martingale property is defined more generally than for integrable functions for usability reasons, but will of course have interesting properties only for integrable functions.

An important example is the martingale generated by a function $f$, defined by $(\mu[f \mid \mathscr F_i])_{i \in \iota}$. 
A consequence of the $L^1$ martingale convergence theorem is that any real uniformly integrable martingale indexed by $\mathbb{N}$ is of that form for some function $f$, as we will see in the following sections.

\section{Martingale convergence theorems}
\label{sec:convergence}
% !TeX root = ../martingales_in_lean.tex

We will in this section describe our formalization of the almost everywhere and the \(L^1\) martingale 
convergence theorems. 

In the remainder of the paper, the time index for stochastic processes will be $\mathbb{N}$ and these processes will be real-valued (although a few results are stated for more general indexes), since Doob's convergence theorems are results about such processes.
The hypotheses of the theorems also use the term \lstinline|measurable| instead of the \lstinline|strongly_measurable| property explained in the introduction, but these two notions coincide for real valued random variables.

\subsection{The almost everywhere martingale convergence theorem} % (fold)
\label{sub:the_ae_convergence_theorem}

We recall the statement of the almost everywhere martingale convergence theorem as mentioned in the introduction.

\begin{theorem}[Almost everywhere martingale convergence] 
  Given a submartingale \((f_n)_{n \in \mathbb{N}}\) adapted to the filtration \((\mathscr{F}_n)_{n \in \mathbb{N}}\), 
  if there exists some \(R \in \mathbb{R}\) such that \(\|f_n\|_1 \le R\) for all \(n \in \mathbb{N}\), then there 
  exists some \(\bigvee_{n \in \mathbb{N}} \mathscr{F}_n\)-measurable function \(f\) which is integrable such that 
  \(f_n \xrightarrow{\text{a.e.}} f\).

  We will for shorthand denote \(\bigvee_{n \in \mathbb{N}} \mathscr{F}_n\) by \(\mathscr{F}_\infty\).
\end{theorem}

The proof of the theorem relies on the following crucial observation on the convergence of real sequences. Namely, 
a real sequence \((x_n)\) converges if and only if 
\begin{enumerate}[(a)]
  \item\label{cond:a} \((|x_n|)\) is bounded above;
  \item\label{cond:b} and for all \(a < b \in \mathbb{Q}\), \((x_n)\) does not visit below \(a\) and above \(b\) infinitely often.
\end{enumerate}
Thus, for a submartingale \((f_n)\) satisfying the assumptions of the theorem, it suffices to show that \((f_n)\) 
satisfies the conditions \ref{cond:a} and \ref{cond:b} almost everywhere. To show this, we will introduce Doob's 
upcrossing estimate which provides a bound on the expected number of times \((f_n)\) can cross from below \(a\) 
to above \(b\).

% subsection the_almost (end)

\subsection{Doob's upcrossing estimate}
\label{sub:upcrossing}

Given real values \(a < b\), we would like to define a notion of ``upcrossings'' of \((f_n)\) 
across the band \([a, b]\) which counts the number of times any realization of \((f_n)\) crosses 
from below \(a\) to above \(b\). To make this heuristic rigorous, we introduce the following stopping 
times mutually inductively defined using hitting times,
\[\sigma_0 := 0, \tau_n := H_{\sigma_n, N}^f(-\infty, a], \text{ and } 
  \sigma_{n + 1} := H_{\tau_{n + 1}, N}^f[b, \infty),\]
for some given \(N \in \mathbb{N}\). Then, we may define the upcrossings of \((f_n)\) before time 
\(N\) to be 
\[U_N^f(a, b) := \sup \{n \mid \sigma_n < N\}.\]

In Lean, the stopping times $(\sigma_n)_{n \in \mathbb{N}}$ and $(\tau_n)_{n \in \mathbb{N}}$ are 
defined as
\begin{lstlisting}
def upper_crossing [preorder ι] [order_bot ι] [has_Inf ι]
  (a b : ℝ) (f : ι → Ω → ℝ) (N : ι) : ℕ → Ω → ι
| 0 := ⊥
| (n + 1) := λ x, hitting f (set.Ici b) 
  (lower_crossing_aux a f (upper_crossing n x) N x) N x
\end{lstlisting}
and 
\begin{lstlisting}
def lower_crossing 
  [preorder ι] [order_bot ι] [has_Inf ι]
  (a b : ℝ) (f : ι → Ω → ℝ) (N : ι) (n : ℕ) : Ω → ι :=
λ x, hitting f (set.Iic a) (upper_crossing a b f N n x) N x
\end{lstlisting}
respectively, where \lstinline{lower_crossing_aux} is defined by
\begin{lstlisting}
def lower_crossing_aux [preorder ι] [has_Inf ι] 
  (a : ℝ) (f : ι → Ω → ℝ) (c N : ι) : Ω → ι :=
hitting f (set.Iic a) c N
\end{lstlisting}

As an example, suppose we have the following realization of a process crossing the band \([a, b]\) in which we take \(N\) to be 
13. Then, by construction, \(\sigma_0 = 0\) and as \(\tau_0\) is the first time the process hits below \(a\), 
\(\tau_0 = 1\). Similarly, as \(\sigma_1\) is the first time the process hits above \(b\) after \(\tau_0\), \(\sigma_1 = 5\). 
By the same rationale, \(\tau_1 = 7\) and \(\sigma_2 = 10\). Finally, as the realization does not cross another band before 
13, \(\sigma_n, \tau_n = 13\) for all \(n > 2\).

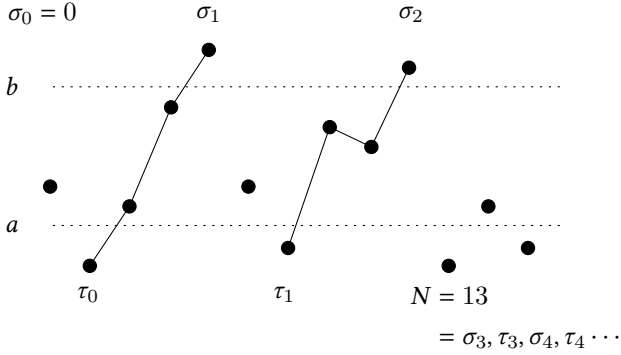
\begin{figure}[H]

\centering
\begin{tikzpicture}[x=0.75pt,y=0.75pt,yscale=-1,xscale=1]

\draw  [dash pattern={on 0.84pt off 2.51pt}]  (25,122) -- (283.43,122) ;
\draw  [dash pattern={on 0.84pt off 2.51pt}]  (25,52) -- (283.43,52) ;

\draw  [fill={rgb, 255:red, 0; green, 0; blue, 0 }  ,fill opacity=1 ] (20.29,102.36) .. controls (20.29,100.5) and (21.79,99) .. (23.64,99) .. controls (25.5,99) and (27,100.5) .. (27,102.36) .. controls (27,104.21) and (25.5,105.71) .. (23.64,105.71) .. controls (21.79,105.71) and (20.29,104.21) .. (20.29,102.36) -- cycle ;
\draw  [fill={rgb, 255:red, 0; green, 0; blue, 0 }  ,fill opacity=1 ] (120.29,102.36) .. controls (120.29,100.5) and (121.79,99) .. (123.64,99) .. controls (125.5,99) and (127,100.5) .. (127,102.36) .. controls (127,104.21) and (125.5,105.71) .. (123.64,105.71) .. controls (121.79,105.71) and (120.29,104.21) .. (120.29,102.36) -- cycle ;
\draw  [fill={rgb, 255:red, 0; green, 0; blue, 0 }  ,fill opacity=1 ] (100.29,33.36) .. controls (100.29,31.5) and (101.79,30) .. (103.64,30) .. controls (105.5,30) and (107,31.5) .. (107,33.36) .. controls (107,35.21) and (105.5,36.71) .. (103.64,36.71) .. controls (101.79,36.71) and (100.29,35.21) .. (100.29,33.36) -- cycle ;
\draw  [fill={rgb, 255:red, 0; green, 0; blue, 0 }  ,fill opacity=1 ] (81.29,62.36) .. controls (81.29,60.5) and (82.79,59) .. (84.64,59) .. controls (86.5,59) and (88,60.5) .. (88,62.36) .. controls (88,64.21) and (86.5,65.71) .. (84.64,65.71) .. controls (82.79,65.71) and (81.29,64.21) .. (81.29,62.36) -- cycle ;
\draw  [fill={rgb, 255:red, 0; green, 0; blue, 0 }  ,fill opacity=1 ] (60.29,112.36) .. controls (60.29,110.5) and (61.79,109) .. (63.64,109) .. controls (65.5,109) and (67,110.5) .. (67,112.36) .. controls (67,114.21) and (65.5,115.71) .. (63.64,115.71) .. controls (61.79,115.71) and (60.29,114.21) .. (60.29,112.36) -- cycle ;
\draw  [fill={rgb, 255:red, 0; green, 0; blue, 0 }  ,fill opacity=1 ] (40.29,142.36) .. controls (40.29,140.5) and (41.79,139) .. (43.64,139) .. controls (45.5,139) and (47,140.5) .. (47,142.36) .. controls (47,144.21) and (45.5,145.71) .. (43.64,145.71) .. controls (41.79,145.71) and (40.29,144.21) .. (40.29,142.36) -- cycle ;
\draw  [fill={rgb, 255:red, 0; green, 0; blue, 0 }  ,fill opacity=1 ] (140.29,133.36) .. controls (140.29,131.5) and (141.79,130) .. (143.64,130) .. controls (145.5,130) and (147,131.5) .. (147,133.36) .. controls (147,135.21) and (145.5,136.71) .. (143.64,136.71) .. controls (141.79,136.71) and (140.29,135.21) .. (140.29,133.36) -- cycle ;
\draw  [fill={rgb, 255:red, 0; green, 0; blue, 0 }  ,fill opacity=1 ] (161.29,72.36) .. controls (161.29,70.5) and (162.79,69) .. (164.64,69) .. controls (166.5,69) and (168,70.5) .. (168,72.36) .. controls (168,74.21) and (166.5,75.71) .. (164.64,75.71) .. controls (162.79,75.71) and (161.29,74.21) .. (161.29,72.36) -- cycle ;
\draw  [fill={rgb, 255:red, 0; green, 0; blue, 0 }  ,fill opacity=1 ] (182.29,82.36) .. controls (182.29,80.5) and (183.79,79) .. (185.64,79) .. controls (187.5,79) and (189,80.5) .. (189,82.36) .. controls (189,84.21) and (187.5,85.71) .. (185.64,85.71) .. controls (183.79,85.71) and (182.29,84.21) .. (182.29,82.36) -- cycle ;
\draw  [fill={rgb, 255:red, 0; green, 0; blue, 0 }  ,fill opacity=1 ] (201.29,42.36) .. controls (201.29,40.5) and (202.79,39) .. (204.64,39) .. controls (206.5,39) and (208,40.5) .. (208,42.36) .. controls (208,44.21) and (206.5,45.71) .. (204.64,45.71) .. controls (202.79,45.71) and (201.29,44.21) .. (201.29,42.36) -- cycle ;
\draw  [fill={rgb, 255:red, 0; green, 0; blue, 0 }  ,fill opacity=1 ] (221.29,142.36) .. controls (221.29,140.5) and (222.79,139) .. (224.64,139) .. controls (226.5,139) and (228,140.5) .. (228,142.36) .. controls (228,144.21) and (226.5,145.71) .. (224.64,145.71) .. controls (222.79,145.71) and (221.29,144.21) .. (221.29,142.36) -- cycle ;
\draw  [fill={rgb, 255:red, 0; green, 0; blue, 0 }  ,fill opacity=1 ] (241.29,112.36) .. controls (241.29,110.5) and (242.79,109) .. (244.64,109) .. controls (246.5,109) and (248,110.5) .. (248,112.36) .. controls (248,114.21) and (246.5,115.71) .. (244.64,115.71) .. controls (242.79,115.71) and (241.29,114.21) .. (241.29,112.36) -- cycle ;
\draw  [fill={rgb, 255:red, 0; green, 0; blue, 0 }  ,fill opacity=1 ] (261.29,133.36) .. controls (261.29,131.5) and (262.79,130) .. (264.64,130) .. controls (266.5,130) and (268,131.5) .. (268,133.36) .. controls (268,135.21) and (266.5,136.71) .. (264.64,136.71) .. controls (262.79,136.71) and (261.29,135.21) .. (261.29,133.36) -- cycle ;

\draw    (43.64,142.36) -- (63.64,112.36) ;
\draw    (63.64,112.36) -- (84.64,62.36) ;
\draw    (84.64,62.36) -- (103.64,33.36) ;
\draw    (143.64,133.36) -- (164.64,72.36) ;
\draw    (164.64,72.36) -- (185.64,82.36) ;
\draw    (185.64,82.36) -- (204.64,42.36) ;

\draw (0,118.4) node [anchor=north west][inner sep=0.75pt]    {$a$};
\draw (0,45.4) node [anchor=north west][inner sep=0.75pt]    {$b$};
\draw (204,151.4) node [anchor=north west][inner sep=0.75pt]    {
  $\begin{aligned}
    N &= 13\\
      &= \sigma_{3}, \tau_{3}, \sigma_{4}, \tau_{4} \cdots 
\end{aligned}$};
\draw (1,8.4) node [anchor=north west][inner sep=0.75pt]    {$\sigma_{0} =0$};
\draw (36,151.4) node [anchor=north west][inner sep=0.75pt]    {$\tau_{0}$};
\draw (96,10.4) node [anchor=north west][inner sep=0.75pt]    {$\sigma_{1}$};
\draw (198,10.4) node [anchor=north west][inner sep=0.75pt]    {$\sigma_{2}$};
\draw (135,151.4) node [anchor=north west][inner sep=0.75pt]    {$\tau_{1}$};

\end{tikzpicture}
\caption{An example of realization of a stochastic process with \(\sigma_n, \tau_n\) labeled.}
\end{figure}

With this, the number of upcrossings $U_N^f(a, b)$ is
\begin{lstlisting}
def upcrossing_before 
  [preorder ι] [order_bot ι] [has_Inf ι]
  (a b : ℝ) (f : ι → α → ℝ) (N : ι) (x : α) : ℕ :=
Sup {n | upper_crossing a b f N n x < N}
\end{lstlisting}
We are allowing arbitrary indices in contrast to just natural number indices as 
we would eventually like to use the same definition for the discretization of continuous martingales.

We may now formulate Doob's upcrossing estimate.

\begin{theorem}[Doob's upcrossing estimate]
  For all \(a < b \in \mathbb{R}\) and \(N \in \mathbb{N}\), we have 
  \begin{equation}\label{eq:doob1}
    (b - a) \mu[U_N^f(a, b)] \le \mu[(f_N - a)^+],
  \end{equation}
  where \(g^+\) denotes the positive part of the function \(g\).
  Furthermore, denoting \(U^f(a, b) := \sup_{N \in \mathbb{N}} U_N^f(a, b)\), we have
  \begin{equation}\label{eq:doob2}
    (b - a) \mu[U^f(a, b)] \le 
      \sup_{N \in \mathbb{N}} \mu[(f_N - a)^+].
  \end{equation}
\end{theorem}

Our formalization closely follows the proof presented in \cite{kallenberg} while making some small adjustments. 
In contrast to proof from \cite{kallenberg} in which the theorem is reduced to the case where \(a = 0\) and \(f_n \ge 0\) for 
all \(n\), we make a weaker reduction where the theorem is reduced to the case where 
\(0 \le f_0\) and \(a \le f_N\) using the equality
\[U_N^{(f - a)^+}(0, b - a) = U_N^f(a, b).\]
In particular, fixing \(a < b \in \mathbb{R}\) and \(N \in \mathbb{N}\), 
it suffices to prove \[(b - a) \mu[U_N^f(a, b)] \le \mu[f_N]\]
in the special case where \(0 \le f_0\) and \(a \le f_N\). Indeed, in the general case, we may 
apply the above inequality to \(((f_n - a)^+)_n\) where \(((f_n - a)^+)_n\) is also a 
submartingale by a simple computation using the monotonicity of the conditional expectation.
The reason for this weaker reduction will be made clear once we have presented the proof.

With the above reduction, by constructing a suitable predictable process satisfying the following lemma, 
the inequality follows by considering the submartingale property of its discrete stochastic integral 
with some predictable process. 

\begin{definition}
  A stochastic process \((c_n)_{n \in \mathbb{N}}\) is said to be predictable with respect to the 
  filtration \((\mathscr{F}_n)_{n \in \mathbb{N}}\) if for all \(n\), \(c_{n + 1}\) is \(\mathscr{F}_n\)-measurable.
\end{definition}

\begin{lemma}
  Given a processes \((f_n)_{n \in \mathbb{N}}\) and \((c_n)_{n \in \mathbb{N}}\), 
  let us denote the discrete stochastic integral of \(c\) with \(f\) the process
  \[(c \cdot f)_n := \sum_{k = 0}^{n - 1} c_{k + 1} (f_{k + 1} - f_k).\]
  Then, if \((f_n)_{n \in \mathbb{N}}\) is a submartingale adapted to the filtration 
  \((\mathscr{F}_n)_{n \in \mathbb{N}}\), and \((c_n)_{n \in \mathbb{N}}\) is 
  predictable, non-negative and bounded, \(((c \cdot f)_n)\) is also a submartingale.
\end{lemma}

We omit the proof of this lemma as it follows by a simple computation.
\begin{lstlisting}
lemma submartingale.sum_mul_sub [is_finite_measure μ] 
  {R : ℝ} {ξ f : ℕ → Ω → ℝ}
  (hf : submartingale f ℱ μ) (hξ : adapted ℱ ξ)
  (hbdd : ∀ n ω, ξ n ω ≤ R) (hnonneg : ∀ n ω, 0 ≤ ξ n ω) :
submartingale 
  (λ n, ∑ k in finset.range n, ξ k * (f (k + 1) - f k)) ℱ μ
\end{lstlisting}

With the above lemma, we will now construct a predictable process satisfying the above lemma and use the 
submartingale property to obtain an inequality bounding the discrete stochastic integral.
Define \(c_n := \sum_{k < N} \mathbf{1}_{[\sigma_k, \tau_{k + 1})}(n)\). It is clear that \((1 - c_n)\) 
is non-negative, bounded and predictable, and hence, \((1 - c) \cdot f\) is again a submartingale. 
Then, by the very definition of a submartingale, 
\(0 \le \mu[((1 - c) \cdot f)_0] \le \mu[((1 - c) \cdot f)_N]\) implying
\[\mu[(c \cdot f)_N] \le \mu[(1 \cdot f)_N] = \mu[f_N] - \mu[f_0]
  \le \mu[f_N].\]
Now, by noting 
\begin{align*}
  (c \cdot f)_N & =
    \sum_{n \le N} \sum_{k \le N} \mathbf{1}_{[\sigma_k, \tau_{k + 1})}(n)(f_{n + 1} - f_n)\\
  & = \sum_{k \le N} \sum_{n \le N} \mathbf{1}_{[\sigma_k, \tau_{k + 1})}(n)(f_{n + 1} - f_n)\\
  & = \sum_{k \le N} (f_{\sigma_k + 1} - f_{\sigma_k} + f_{\sigma_k + 2} - \cdots + 
    f_{\tau_{k + 1}} - f_{\tau_{k + 1} - 1})\\
  & = \sum_{k \le N} (f_{\tau_{k + 1}} - f_{\sigma_k}) \ge \sum_{k < U_N(a, b)} (b - a)\\
  & = (b - a) U_N(a, b)
\end{align*}
where the inequality follows since 
\begin{itemize}
  \item for \(k < U_N(a, b)\), \(f_{\tau_{k + 1}} - f_{\sigma_k} \ge b - a\);
  \item for \(k > U_N(a, b)\), \(f_{\tau_{k + 1}} = f_{\sigma_k} = f_N\) implying 
    \[f_{\tau_{k + 1}} - f_{\sigma_k} = 0;\] 
  \item for \(k = U_N(a, b)\), 
    \[f_{\tau_{k + 1}} - f_{\sigma_k} = f_{\tau_{U_N(a, b) + 1}} - f_{\sigma_{U_N(a, b)}} = f_N - a \ge 0.\]
\end{itemize}
We have
\begin{align*}
  (b - a) \mu[U_N(a, b)] & \le \mu[(c \cdot f)_N] \le \mu[f_N],
\end{align*}
as required.

The overall proof strategy is conducted as described above with a large part of the formalization 
effort spent building a sufficiently large API for the definitions 
\lstinline{upper_crossing} and \lstinline{lower_crossing}. Indeed, due to their definitions, many 
seemingly obvious statements regarding \((\sigma_n)\) and \((\tau_n)\) often involve long and tedious 
mutual inductions. As an example, while the equality 
\[U_N^{(f - a)^+}(0, b - a) = U_N^f(a, b),\]
is rather intuitive: the number of upcrossings is invariant under translations and reflections provided we also 
translate the bands by the same amount; the formalization of this fact spanned almost 100 lines of code.
As a result, we sought to minimize proof steps requiring properties of \((\sigma_n)\) and \((\tau_n)\)
resulting in the aforementioned weaker reduction compared to that is presented in \cite{kallenberg}.

With this, the first inequality of Doob' upcrossing estimate is formalized following the above 
outlined proof in Lean as 
\begin{lstlisting}
lemma upcrossing_estimate [is_finite_measure μ]
  (a b : ℝ) (hf : submartingale f ℱ μ) (N : ℕ) :
  (b - a) * μ[upcrossing_before a b f N] ≤ μ[λ x, (f N x - a)⁺]
\end{lstlisting} 

We note that in contrast to the theorem statement, our formulation does not require \(a < b\). 
Indeed, this assumption is not necessarily as in the case \(a \ge b\), \(b - a < 0\) implying 
the left hand side is non-positive while the right hand side is non-negative.
This remark may appear inconsequential, but it means that we dispense the user of that lemma from having to prove trivial inequalities like $a < b$, which is very desirable in a formal library.

To formalize inequality~(\ref{eq:doob2}) of Doob's upcrossing estimate, we will introduce a new definition 
known as \lstinline{upcrossings}. While we have defined \(U_N^f(a, b)\) to take value in the natural numbers, 
\(U^f(a, b)\) cannot, as it is not necessarily finite since \((U_N^f(a, b))_N\) might not be 
bounded above in \(N\). In the case that \((U_N^f(a, b))_N\) is not bounded above, we say on paper that 
\[\sup_N U_N^f(a, b) = \infty.\]
This does not translate directly into Lean as the supremum of an unbounded set of natural numbers 
is defined to be 0. As a result, to define \(U^f(a, b)\) so that \(U^f(a, b) = \infty\) 
makes sense, we need to first coerce \(U_N^f(a, b)\) to take value in the extended non-negative 
reals \(\mathbb{R}_{\ge 0} \cup \{\infty\}\). 
\begin{lstlisting}
  def upcrossings [preorder ι] [order_bot ι] [has_Inf ι]
  (a b : ℝ) (f : ι → Ω → ℝ) (ω : Ω) : ℝ≥0∞ :=
  ⨆ N, (upcrossings_before a b f N ω : ℝ≥0∞)
\end{lstlisting}
We note that \(U_N^f(a, b)\) is coerced to 
\(\mathbb{R}_{\ge 0} \cup \{\infty\}\) rather than \(\mathbb{N} \cup \{\infty\}\) since as we intend to 
integrate \(U_N^f(a, b)\), if \(U_N^f(a, b)\) is coerced into \(\mathbb{N} \cup \{\infty\}\), the integration 
will introduce another coercion into \(\mathbb{R}_{\ge 0} \cup \{\infty\}\), making the original coercion 
redundant.

With this definition, we may now phrase inequality~(\ref{eq:doob2}) as
\begin{lstlisting}
lemma mul_lintegral_upcrossings_le_lintegral_pos_part 
  [is_finite_measure μ] (a b : ℝ) (hf : submartingale f ℱ μ) :
  ennreal.of_real (b - a) * ∫⁻ ω, upcrossings a b f ω ∂μ ≤
    ⨆ N, ∫⁻ ω, ennreal.of_real ((f N ω - a)⁺) ∂μ
\end{lstlisting}

In contrast to the first inequality in Doob's upcrossing estimate where the expectation is formulated 
using the Bochner integral, as \lstinline{upcrossings} takes value in the non-negative extended 
reals, the target space is no longer a Banach space and so we may not use the Bochner integral. Instead, 
the expectation is formulated using the lower Lebesgue integral (known as \lstinline{lintegral} in Lean).

The proof of inequality~(\ref{eq:doob2}) is rather simple. Indeed, by taking the supremum of both sides
of inequality~(\ref{eq:doob1}), inequality~(\ref{eq:doob2}) follows by the monotone convergence theorem 
where we notice \((U_N^f(a, b))_N\) is monotone in \(N\). However, it is important to note that, to 
apply the monotone convergence theorem, we need to make sure \((U_N^f(a, b))_N\) is both measurable 
and integrable for all \(N\). Integrability is easy to check as \(U^f_N(a, b)\) is bounded above by \(N\), 
and so it suffices to show measurability. Indeed, by noting that
\[U^f_N(a, b) = \sum_{i = 1}^N \mathbf{1}_{\{n \in \mathbb{N} \mid \sigma_{n} < N\}}(i)\]
\(U^f_N(a, b)\) is measurable as \(\{n \mid \sigma_{n} < N\}\) is a measurable set since \((\sigma_n)\) is a 
stopping time.

With the Doob's upcrossing estimate, we may now finish the proof of the almost everywhere martingale 
convergence theorem.
We recall that a real sequence \((x_n)\) converges if and only if 
\begin{enumerate}[(a)]
  \item\label{cond:a2} \((|x_n|)\) is bounded above;
  \item\label{cond:b2} and for all \(a < b \in \mathbb{Q}\), \((x_n)\) does not visit below \(a\) and above \(b\) infinitely often.
\end{enumerate}
Thus, for a submartingale \((f_n)\) adapted to the filtration \((\mathscr{F}_n)\) such that for all \(n\), 
\(\|f_n\|_1 \le R\) for some \(R \in \mathbb{R}\), we will show that \((f_n)\) satisfies 
the above two conditions almost everywhere. 

Firstly, as \((f_n)\) is uniformly bounded in \(L^1\) by \(R\), by inequality~(\ref{eq:doob2}) of 
Doob's upcrossing estimate, we have 
\[(b - a) \mu[U^f(a, b)] \le \sup_{N \in \mathbb{N}} \mu[(f_N - a)^+]
  \le R + |a| \cdot \mu(\Omega),\]
implying \(\mu[U^f(a, b)]\) is finite and so, \(U^f(a, b)\) is finite almost everywhere. 
Thus, as \(U^f(a, b) = \infty\) if and only if \((f_n)\) does not visit below \(a\) and above \(b\) infinitely 
often, \((f_n)\) satisfies condition~\ref{cond:b} almost everywhere. 

On the other hand, \((|f_n|)\) is bounded almost everywhere. Indeed, in the case that \((|f_n(\omega)|)\) is not 
bounded above, either 
\begin{itemize}
  \item one of \(\liminf_{n} f_n(\omega)\) or 
    \(\limsup_{n} f_n(\omega)\) equals \(-\infty, +\infty\) respectively with the other being finite,
  \item or \(f_n(\omega) \to \pm \infty\).
\end{itemize}
The first case is a subset of \(\{U^f(a, b) = \infty\}\) and so can only happen with measure 0  
as shown above. The second case is a subset of \(\{\liminf_{n} |f_n| = \infty\}\). By Fatou's lemma, we have
\[\mu [\limsup_{n \to \infty} |f_n|] \le \limsup_{n \to \infty} \mu [|f_n|] < \infty\]
implying \(\liminf_{n} |f_n| < \infty\) almost everywhere and so, the second case can also only 
happen with measure 0. 

With this, we conclude that \((f_n)\) converges almost everywhere and it remains to show that it has an a.e. 
limit which is integrable and measurable with respect to \(\mathscr{F}_\infty\). The definition of a.e. measurable 
becomes especially useful here. Traditionally, one obtains measurability by constructing the a.e. limit to be 
\(\liminf_n f_n\) which is \(\mathscr{F}_\infty\)-measurable. Instead, as an a.e. pointwise 
limit of a sequence of measurable functions is at least a.e. measurable, since \(f_n\) is 
\(\mathscr{F}_\infty\)-measurable for all \(n\), if \(f_n \xrightarrow{a.e.} f\) then \(f\) is 
\(\mathscr{F}_\infty\)-a.e. measurable, namely, there exists some \(\mathscr{F}_\infty\)-measurable 
\(f'\) equal to \(f\) almost everywhere. Thus, \(f_n \xrightarrow{a.e.} f'\) and hence, choosing 
our limiting function to be \(f'\) suffices. 

Finally, integrability of the limiting function is clear by considering the following corollary of 
Fatou's lemma.

\begin{lemma}
  If \((f_n)\) is a sequence of measurable functions such that \(f_n \xrightarrow{a.e.} f\), then 
  \[\|f\|_p \le \liminf_{n \to \infty} \|f_n\|_p,\]
  for all \(p \in \mathbb{R}_{\ge 0} \cup \{\infty\}\).
\end{lemma}

We note that our proof is slightly more roundabout than the usual proof found in literature where we 
conclude the almost everywhere convergence directly after noticing that a real sequence \((x_n)\) converges if and only if 
\(\liminf_n |x_n| < \infty\) and for all \(a < b \in \mathbb{Q}\), \((x_n)\) crosses the band \([a, b]\) 
only finitely many times. The reason choosing the indirect route is that
the key observation in our argument where we characterize the convergence of real sequences was independently 
proven by Gouëzel for the purpose of proving the Lebesgue differentiation theorem \cite{gouezel}. Thus, by 
changing our proof to utilize Gouëzel's work, we are able to avoid redundant work.

\paragraph{Defining a limit function instead of stating existence}

Our formalization of the a.e. martingale convergence theorem follows the proof outlined above with one 
notable implementation difference. While we had stated the a.e. martingale convergence theorem as the 
existence of a limiting random variable satisfying some properties, since applications of the theorem 
often require us to manipulate this limit, it is easier to simply extract the limit of the process into 
its own definition. In particular, we define the \lstinline{limit_process} of a sequence of 
random variables \((f_n)\) with respect to the filtration \((\mathscr{F}_n)\) and the measure \(\mu\) 
as the \(\mathscr{F}_\infty\)-measurable, \(\mu\)-a.e. limit of \((f_n)\) if it exists, otherwise 
we define it as 0.
\begin{lstlisting}
def limit_process (f : ι → Ω → E) (ℱ : filtration ι 𝒜) 
  (μ : measure Ω) :=
if h : ∃ g : Ω → E, strongly_measurable[⨆ n, ℱ n] g ∧
  ∀ᵐ ω ∂μ, tendsto (λ n, f n ω) at_top (𝓝 (g ω)) 
  then classical.some h else 0
\end{lstlisting}
Here, using the axiom of choice, \lstinline{classical.some h} for \lstinline{h : ∃ x : α, p x} 
where \lstinline{p} is a predicate on the type \lstinline{α} provides a term of \lstinline{α} which 
satisfies \lstinline{p}.

\lstinline{limit_process} is a useful definition. By extracting the definition, not only do we avoid 
needing to deconstruct the a.e. martingale convergence theorem every time we need to manipulate the 
limiting random variable, we can also directly reason about the definition itself. This is 
exemplified by the following lemmas in which \lstinline{f} is not required to be a submartingale. 
\begin{lstlisting}
lemma strongly_measurable_limit_process :
  strongly_measurable[⨆ n, ℱ n] (limit_process f ℱ μ)
\end{lstlisting}
\begin{lstlisting}
lemma mem_ℒp_limit_process_of_snorm_bdd 
  (hfm : ∀ n, ae_strongly_measurable (f n) μ) 
  (hbdd : ∀ n, snorm (f n) p μ ≤ R) :
  mem_ℒp (limit_process f ℱ μ) p μ
\end{lstlisting}
With this definition, the a.e. martingale convergence theorem is phrased and formalized as 
\begin{lstlisting}
lemma submartingale.ae_tendsto_limit_process 
  [is_finite_measure μ] (hf : submartingale f ℱ μ) 
  (hbdd : ∀ n, snorm (f n) 1 μ ≤ R) :
  ∀ᵐ ω ∂μ, tendsto (λ n, f n ω) at_top 
    (𝓝 (ℱ.limit_process f μ ω))
\end{lstlisting}

\subsection{\(L^1\) martingale convergence theorem}

Continuing, we now formulate and prove the \(L^1\) martingale convergence theorem. 
The \(L^1\) martingale convergence theorem is commonly phrased as the following equivalence.
\begin{theorem}[\(L^1\) martingale convergence]
  Given \((f_n)_{n \in \mathbb{N}}\) a martingale adapted to the filtration \((\mathscr{F}_n)_{n \in \mathbb{N}}\), 
  the following are equivalent:
  \begin{enumerate}[(i)]
    \item\label{cond:i} \((f_n)\) is uniformly integrable (in the probability sense);
    \item\label{cond:ii} \((f_n)\) converges in \(L^1\) to some \(\mathscr{F}_\infty\)-measurable, integrable function \(f\);
    \item\label{cond:iii} there exists some \(\mathscr{F}_\infty\)-measurable and integrable \(f\) such that for all \(n \in \mathbb{N}\), 
    $\mu[f \mid \mathscr{F}_n] = f_n \text{ a.e.}$ .
  \end{enumerate}
\end{theorem}
However, while this theorem is stated for martingales, several implications from the proof of this theorem also hold for submartingales.
So, rather than proving the theorem as stated above, we extract these implications into their own statements and shall instead consider the following slightly more general theorem.
\begin{theorem}\label{l1-conv}
  Given some filtration \((\mathscr{F}_n)_{n \in \mathbb{N}}\), 
  \begin{enumerate}[(a)]
    \item\label{res:a} if \((f_n)_{n \in \mathbb{N}}\) is a submartingale adapted to \((\mathscr{F}_n)\) and is 
      uniformly integrable in the probability sense, it converges in \(L^1\) to some 
      \(\mathscr{F}_\infty\)-measurable, integrable function \(f\);
    \item\label{res:b} if furthermore, \((f_n)\) is a martingale, then for all \(n \in \mathbb{N}\),
    \[\mu[f \mid \mathscr{F}_n] = f_n \text{ a.e.} \: ;\]
    \item\label{res:c} finally, if \(g\) is some integrable, \(\mathscr{F}_\infty\)-measurable function, then 
    \((\mu[g \mid \mathscr{F}_n])_{n \in \mathbb{N}}\) converges to \(g\) a.e. and in \(L^1\).
  \end{enumerate}
\end{theorem}
Indeed, these three statements implies the \(L^1\) martingale convergence theorem directly where the implication 
\ref{cond:i} \(\implies\) \ref{cond:ii} follows from \ref{res:a}, the implication 
\ref{cond:ii} \(\implies\) \ref{cond:iii} follows from \ref{res:b}, and the implication 
\ref{cond:iii} \(\implies\) \ref{cond:i} follows from \ref{res:c}.

As we have already constructed a sufficiently large amount of API for uniform integrability, 
the proof of theorem~\ref{l1-conv} is rather simple. By recalling that uniformly integrable 
in the probability sense implies uniform boundedness in \(L^1\), \ref{res:a} follows. Indeed, 
as \((f_n)\) is bounded in \(L^1\), by the a.e. martingale convergence theorem we have \((f_n)\) 
converges almost everywhere to some \(\mathscr{F}_\infty\)-measurable, integrable function \(f\).
Thus, as convergence a.e. implies convergence in measure, we have by the Vitali convergence theorem that 
\((f_n)\) converges in \(L^1\) to \(f\) as required. 

Now, taking \((f_n)\) to be a martingale and denoting its \(L^1\) limit by \(f\) (which exists by \ref{res:a}), 
\ref{res:b} follows since for all \(n\), we have for all \(m \ge n\), 
\[\|f_n - \mu[f \mid \mathscr{F}_n]\|_1 =
\|\mu[f_m - f \mid \mathscr{F}_n]\|_1 \le \|f_m - f\|_1.\]
Thus, \(\|f_n - \mu[f \mid \mathscr{F}_n]\|_1 = 0\) by taking \(m \to \infty\) implying 
\(f_n = \mu[f \mid \mathscr{F}_n]\) a.e.

Finally, we notice on one hand that any class of the form \(\{\mu[g \mid \mathscr{G}_i] \mid i \in \iota\}\) for some 
\(\iota\)-indexed sub-\(\sigma\)-algebras \((\mathscr{G}_i)_{i \in \iota}\) is uniformly integrable, and on the other hand that 
\((\mu[g \mid \mathscr{F}_n])_{n \in \mathbb{N}}\) is a martingale, hence by \ref{res:b} it converges a.e. and in \(L^1\)
to some function \(g'\).
Furthermore, \(\mu[g \mid \mathscr{F}_n] = \mu[g' \mid \mathscr{F}_n]\) almost 
everywhere for all \(n\). Thus, by the law of iterative expectation, for all \(A \in \mathscr{F}_n\), we have
\[\mu[g \mathbf{1}_A] = \mu[\mu[g \mid \mathscr{F}_n]\mathbf{1}_A] = 
  \mu[\mu[g' \mid \mathscr{F}_n] \mathbf{1}_A] = \mu[g' \mathbf{1}_A]\]
for all \(n\). Hence, as \(\bigcup_n \mathscr{F}_n\) is a \(\pi\)-system generating \(\mathscr{F}_\infty\), by 
Dynkin's theorem, the equality also holds for all \(A \in \mathscr{F}_\infty\). Thus, as both \(g, g'\) are 
\(\mathscr{F}_\infty\) measurable, we conclude that \(g = g'\) almost everywhere and so, 
\(\mu[g \mid \mathscr{F}_n] \to g\) a.e. and in \(L^1\) as required.

This theorem in Lean is stated and proved as described above in three parts. Similar to the statement of the a.e. martingale convergence
theorem, instead of showing the existence of an \(L^1\)-limit, by noting that the \(L^1\) limit coincides with the a.e. 
limit, we may alternatively phrase the theorem using \lstinline{limit_process}. 
\begin{lstlisting}
  lemma submartingale.tendsto_snorm_one_limit_process
  (hf : submartingale f ℱ μ) 
  (hunif : uniform_integrable f 1 μ) :
  tendsto (λ n, snorm (f n - ℱ.limit_process f μ) 1 μ) 
  at_top (𝓝 0)
  
  lemma martingale.ae_eq_condexp_limit_process
  (hf : martingale f ℱ μ) 
  (hbdd : uniform_integrable f 1 μ) (n : ℕ) :
  f n =ᵐ[μ] μ[ℱ.limit_process f μ | ℱ n]
  
  lemma integrable.tendsto_snorm_condexp
  (hg : integrable g μ) 
  (hgmeas : strongly_measurable[⨆ n, ℱ n] g) :
  tendsto (λ n, snorm (μ[g | ℱ n] - g) 1 μ) at_top (𝓝 0)
\end{lstlisting}
As the proof of this theorem only uses standard analysis arguments, the formalization was straightforward 
with the help of existing lemmas in \lstinline|mathlib|. 

\subsection{Lévy's generalization of the Borel-Cantelli lemmas}

The martingale convergence theorems has numerous applications in probability theory and provides 
alternative proofs to the SLLN, the Radon-Nikodym theorem, the Borel-Cantelli lemmas etc. While the 
SLLN and the Radon-Nikodym theorem have already been formalized in \lstinline|mathlib| via other methods, 
the second Borel-Cantelli lemma is yet to be formalized. Thus, as an application the martingale 
convergence theorems, we formalize a more general version of the second Borel-Cantelli lemma known 
as Lévy's generalized Borel-Cantelli lemma.

\begin{theorem}[Lévy's generalized Borel-Cantelli lemma]
  Given a filtration \((\mathscr{F}_n)_{n \in \mathbb{N}}\) and a sequence of sets \((S_n)_{n \in \mathbb{N}}\)
  such that \(S_n \in \mathscr{F}_n\) for all \(n\), then 
  \[\limsup_{n \to \infty} S_n = \left\{\sum_{n = 0}^\infty \mu[\mathbf{1}_{S_{n + 1}} \mid \mathscr{F}_n] = \infty\right\}\]
  almost everywhere.
\end{theorem}

This is formalized in Lean as follows:
\begin{lstlisting}
theorem ae_mem_limsup_at_top_iff [is_finite_measure μ]
  {s : ℕ → set Ω} (hs : ∀ n, measurable_set[ℱ n] (s n)) :
  ∀ᵐ ω ∂μ, ω ∈ limsup at_top s ↔
    tendsto (λ n, ∑ k in finset.range n, 
      μ[(s (k + 1)).indicator (1 : Ω → ℝ) | ℱ k] ω)
      at_top at_top
\end{lstlisting}

We shall omit the details of the proof here and refer the reader to \cite{kallenberg}. The proof of this theorem  
illustrated the usage Doob's decomposition theorem to construct martingales. In particular, to prove Lévy's 
generalized Borel-Cantelli lemma, we need to construct the martingale defined by 
\[f_n := \sum_{k < n} \left(\mathbf{1}_{S_{k + 1}} - \mu[\mathbf{1}_{S_{k + 1}} \mid \mathscr{F}_k]\right).\]
It is easy to check that this is indeed a martingale by using the law of iterated expectations.
However, $f_n$ is in fact the martingale part from Doob's decomposition of the process 
\[\tilde f_n := \sum_{k < n} \mathbf{1}_{S_{k + 1}}.\]
Hence, rather than checking \((f_n)\) is a martingale explicitly and since that notion is more generally useful, we chose to implement Doob's decomposition of a process into a martingale part and a predictable part.
To complete our proof, it then suffices to check that \((f_n)\) equals the martingale part of some adapted process \((\tilde f_n)\). 

We remark that an alternative approach is to prove that a process of the form 
\(f_n := \sum_{k \le n} g_k\) is a martingale if \((g_n)\) is an adapted process and 
\(\mu[g_n] = 0\) for all \(n\).

\section{Concluding thoughts}
\label{sec:future}
% !TeX root = ../martingales_in_lean.tex

We added to \lstinline|mathlib| definitions of the conditional expectation, filtrations, stopping times and martingales and we formalized Doob's convergence theorems about those objects. Formalizing theorems about the new definitions was essential, since it is very hard to know if a definition will be adequate until we actually use it, as exemplified by our change of measurability conditions for \lstinline|uniform_integrable|.
The analysis and measure theory sections of \lstinline|mathlib| have now matured enough for probability theory to be developed and we see no obstacle to the formalization of other fundamental tools of probability.
We saw with the betting example in the introduction that martingales can be used to model the interaction of an player with a stochastic environment. Suppose now that the player can choose among several coins which might be differently biased, and makes that decision adaptively based on the result of the previously observed coin flips\footnote{that interaction is called a stochastic bandit \cite{lattimore2020bandit}}. Then the distribution of the next reward is random since it depends on past observations. To model such interactions we need Markov kernels and regular conditional probability, which we see as a natural next step in our formalization effort.

\paragraph{Retrospection of a design decision}

The formalization of Lévy's generalized Borel-Cantelli illustrates an annoyance with our current formulation of the theory of 
stopping times. Our current formulation of stopping times is similar to the Isabelle/HOL \cite{probzoo} 
formulation in the sense that we let the range of a stopping time to be the same type as the time index of 
its associated filtration: if the filtration is indexed by $\mathbb{N}$, the stopping time must be finite.
As demonstrated in the definition of hitting times, 
this is slightly different to what we write on paper where it is allowed to be infinity. The 
same is true for stopping times. Indeed, for a stopping time \(\sigma\), on paper we normally 
allow \(\sigma\) to take value infinity and then work with \(\sigma \wedge n\). This is not 
possible with our current set up. This discrepancy is not problematic in our formalization of the martingale 
convergence theorems, however we observe that it introduces code duplication in the formalization of the 
Lévy's generalized Borel-Cantelli. In particular, Lévy's generalized Borel-Cantelli requires a lemma known as 
the one-sided martingale bound which proof starts by
introducing a stopping time \(\sigma\) which can take value infinity and argue that the stopped process 
\((f_{\sigma \wedge n})\) of a submartingale \((f_n)\) is also a submartingale.
We however cannot proceed like this since we cannot have a stopping time with possibly infinite value.
Instead, we introduce an object which is almost a stopped process but not exactly since we don't have a stopping time, and 
again reprove a few lemmas specifically for it. A possible way to avoid this duplication is to change the 
definition of stopping time to allow for infinity in the range, namely change the definition to 
\begin{lstlisting}
def is_stopping_time [preorder ι] (ℱ : filtration ι 𝒜) 
  (τ : Ω → with_top ι) :=
∀ i, measurable_set[ℱ i] {ω | τ ω ≤ i}
\end{lstlisting}
\lstinline|with_top ι| is the type \lstinline|ι| together with an infinity value.
Not having the same type for the filtration index and the stopping time range could however be cumbersome since we then can't directly index by a stopping time, and only our upcoming refactor attempt will tell which of the definitions is easier to use.

\section*{Acknowledgements}

The authors would like to thank Sébastien Gouëzel and Kalle Kytölä for their help and suggestions during the formalization process.
They also thank the other \lstinline|mathlib| reviewers and contributors shaping \lstinline|mathlib| without which this project would not 
have been possible. They would also like to thank the conference reviewers for their helpful comments.

\bibliographystyle{ACM-Reference-Format}
\bibliography{biblio}

\end{document}